\documentclass[preprint,12pt]{elsarticle}

\usepackage{subcaption,graphicx}
\usepackage{dcolumn}
\usepackage{bm}
\usepackage{amsfonts}
\usepackage{gensymb}
\usepackage{amsmath}
\usepackage{amssymb}
\usepackage{latexsym}
\usepackage{textcomp}
\usepackage{color,soul}
\usepackage{tikz}
\biboptions{numbers,sort&compress}

\PassOptionsToPackage{normalem}{ulem}
\usepackage{ulem}
\providecolor{added}{rgb}{0,0,1}
\providecolor{deleted}{rgb}{1,0,0}

\journal{Applied Surface Science}

\begin{document}
\def\myfrac#1#2{\frac{\displaystyle #1}{\displaystyle #2}}
\begin{frontmatter}

\title{Femtosecond laser patterning of graphene electrodes for thin-film transistors} 
\author{Maren Kasischke}
\ead{kasischke@lat.rub.de}
\address{Ruhr University Bochum, Faculty of Mechanical Engineering, Applied Laser Technologies, Bochum, Germany}
\author{Ersoy Suba\c{s}{\i}}
\address{Ruhr University Bochum, Faculty of Electrical Engineering and Information Technology, Electronic Materials and Nanoelectronics, Bochum, Germany}
\author{Claudia Bock}
\address{Ruhr University Bochum, Faculty of Electrical Engineering and Information Technology, Microsystems Technology, Bochum, Germany}
\author{Duy-Vu Pham}
\address{Evonik Resource Efficiency GmbH, Electronic Solutions, Marl Germany}
\author{Evgeny L. Gurevich}
\address{Ruhr University Bochum, Faculty of Mechanical Engineering, Applied Laser Technologies, Bochum, Germany}
\author{Ulrich Kunze}
\address{Ruhr University Bochum, Faculty of Electrical Engineering and Information Technology, Electronic Materials and Nanoelectronics, Bochum, Germany}
\author{Andreas Ostendorf}
\address{Ruhr University Bochum, Faculty of Mechanical Engineering, Applied Laser Technologies, Bochum, Germany}
\date{\today}

\begin{abstract}

The aim of this study is to assess femtosecond laser patterning of graphene in air and in vacuum for the application as source and drain electrodes in thin-film transistors (TFTs). The analysis of the laser-patterned graphene with scanning electron microscopy, atomic force microscopy and Raman spectroscopy showed that processing in vacuum leads to less debris formation and thus re-deposited carbonaceous material on the sample compared to laser processing in air. It was found that the debris reduction due to patterning in vacuum improves the TFT characteristics significantly. Hysteresis disappears, the mobility is enhanced by an order of magnitude and the subthreshold swing is reduced from $S_{sub}~=~2.5~\mbox{V/dec}$ to $S_{sub}~=~1.5~\mbox{V/dec}$. 

\end{abstract}

\end{frontmatter}
\section{Introduction}

Graphene is a monolayer of carbon atoms arranged in a honeycomb lattice and presents extraordinary properties such as wavelength-independent absorption in the UV-visible range of $2.3\,\%$ \cite{Nair.2008}, high electrical conductivity due to massless electrons \cite{Novoselov.2005} and high mechanical strength \cite{lee.2008b}. This 2D material has a great potential for electronic devices \cite{Novoselov.2012}. Much research effort has been put into a high quality and up-scalable production \cite{li.2009,Bae.2010} and feasibility of application of graphene \cite{Geim.2009,kim.2009} in the last years. 
The handling and processing of graphene, however, still represents a great challenge, since several processing steps induce an unwanted modification of the material properties. For example, the use of photoresists or polymer films on graphene for photolithography and/or transfer processes cause chemical doping of the layer \cite{Ishigami.2007,Kumar.2011,Kumar.2013,Suk.2013,chavarin.2016} or patterning with e-beam lithography hydrogenates the graphene basal plane causing defects \cite{ryu.2011}. The necessary post-processing cleaning step in Ar/H$_2$ at $250-400~^{\circ}\mbox{C}$ for removing polymer residues can also lead to graphene degradation \cite{Mahmood.2015}.

In order to use graphene in electronic devices or to study its electrical properties it is usually patterned by lithography. Different methods exist, as for example plasma etching \cite{childres.2011,al.2014}, resist-free soft lithography \cite{george.2013}, stencil mask lithography \cite{Mahmood.2015} and helium ion lithography \cite{bell.2009}.

One attractive possibility to pattern graphene is with ultrafast lasers \cite{Kalita.2011,Stohr.2011,Zhang.2012b,Sahin.2014,vanErps.2015,Dong.2016,bobrinetskiy.2016}. Additionally to common benefits of using ultrafast lasers for patterning, like minimized heat affected zone, flexibility and selectivity, this method is especially beneficial for graphene patterning. Indeed, being contactless and requiring no sample preparation, it reduces undesired modification of the graphene. This method can be applied for patterning graphene also on sensitive flexible substrates due to its limited thermal influence. The problem however is that during laser patterning graphene forms folds \cite{Yoo.2012} and debris causing disturbance resulting in unwanted defects and negative impact on its application as transistor electrodes. 
Femtosecond laser ablation of graphene in vacuum, compared to air, should result in cleaner and more controlled patterning of the layer because rather less material is redeposited on the substrate \cite{chichkov.1996,matsumura.2005,feng.2015}. Moreover, atmospheric oxygen can oxidize the remaining graphene upon laser ablation and reduce performance of graphene-based electronics.

This article presents graphene ablation on silicon substrates with a femtosecond laser in air and in vacuum and assesses, if the resulting patterns are appropriate for its application as electrodes in metal-oxide thin-film transistors (MOTFTs). The patterned structures are analyzed topographically, with scanning electron microscopy (SEM) and atomic force microscopy (AFM) and additionally by Raman spectroscopy. After measuring the sheet resistance of the patterned graphene electrodes, these are used as source and drain electrodes in MOTFTs.  

\section{Experimental}
The graphene samples (Graphenea Inc., Spain) are grown by chemical vapor deposition on copper and subsequently transferred onto silicon wafer with thermally grown silicon oxide layer of 300 nm. For laser ablation experiments of graphene the samples are used `as received'. Laser processing is carried out with a fiber-rod amplified femtosecond laser (\textit{Tangerine}, maximal average power $P=20\,\mbox{W}$, wavelength $\lambda=1030\,\mbox{nm}$, repetition rate $f=2\,\mbox{MHz}$, pulse duration $\tau_p\approx 280\,\mbox{fs}$, Amplitude Systemes, France), which is guided and focused onto the sample with a galvanometer scanner and a 63 mm f-theta objective. Taking $M\textsuperscript{2}= 1.05\,$ into account the calculated radius at beam waist is $\omega_0 = 16.8\,\micro\text{m}$. The graphene electrode itself remains unexposed to the laser radiation during the patterning, but the rest of the graphene layer is removed by the laser with a fluence of $F=79~\mbox{mJ/cm}^2$.
The laser power is controlled by an assembly of a $\lambda \backslash 2$ plate and a polarizing beam splitter. For experiments in vacuum the samples are placed in a chamber where a vacuum of $p\approx 1\times10^{-1}\,\mbox{mbar}$ is reached. 

\begin{figure}[h]
\centering
\includegraphics[width=13 cm]{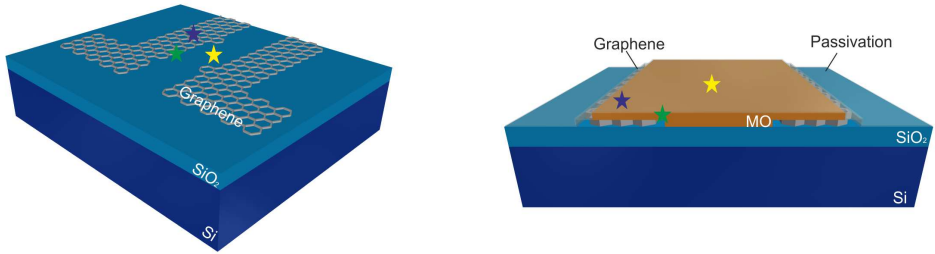}
\caption{Schematic figure of the graphene-based transistor. Left: graphene electrodes patterned by femtosecond laser on top of a thermally oxidized silicon substrate. Right: bottom-gate bottom-contact metal-oxide thin-film transistor with graphene electrodes. The colored stars mark the positions which are analyzed by SEM, AFM and Raman spectroscopy.}
\label{setup}
\end{figure}

Transistors are prepared in bottom-gate bottom-contact configuration (see Fig. \ref{setup} right). After laser-induced patterning of source and drain graphene electrodes Ti/Au bond-pads are e-beam evaporated using a laser-cut shadow mask. The indium based metal-oxide precursor (iXsenic\textsuperscript{\textregistered} S, Evonik Industries AG, Germany) is spin-coated and annealed for one hour at $350\,^{\circ}\mbox{C}$ under atmospheric conditions. The resulting indium oxide layer is referred to as MO (metal oxide) in this article. Finally, a surface passivation layer (iXsenic\textsuperscript{\textregistered} P,  Evonik Industries AG, Germany) was spin-coated, dried, crosslinked in an additional UV-ozone treatment and converted at $350\,^{\circ}\mbox{C}$.

Graphene samples are analyzed with scanning electron microscopy (SEM, Leo Gemini 982, Carl Zeiss AG, Germany) using an acceleration voltage of $1\,\mbox{kV}$ for uncoated graphene and $3\,\mbox{kV}$ for graphene samples with MO and passivation layer coated on top. Additionally, atomic force microscopy (AFM, Nanoscope 5, Bruker Corp., USA) is performed in PeakForce Tapping mode in ScanAsyst mode. Raman measurements (inVia, Renishaw GmbH, Germany) are done with an $100\,\times$ objective, $\lambda=532\,\mbox{nm}$ and less than $0.7~\mbox{mW}$ power in order to avoid damage of the graphene layer. Raman mapping is performed with $1\,\micro \mbox{m}$ step. Acquired spectra are baseline subtracted and intensity ratios of main peaks are plotted as Raman map. In this study the I(D)/I(G) ratio and I(2D)/I(G) ratio are calculated and plotted for each measured spot. Shading of Raman maps was fit by interpolation.

\section{Results}
\begin{figure}[h!]
\centering
\includegraphics[width=12cm]{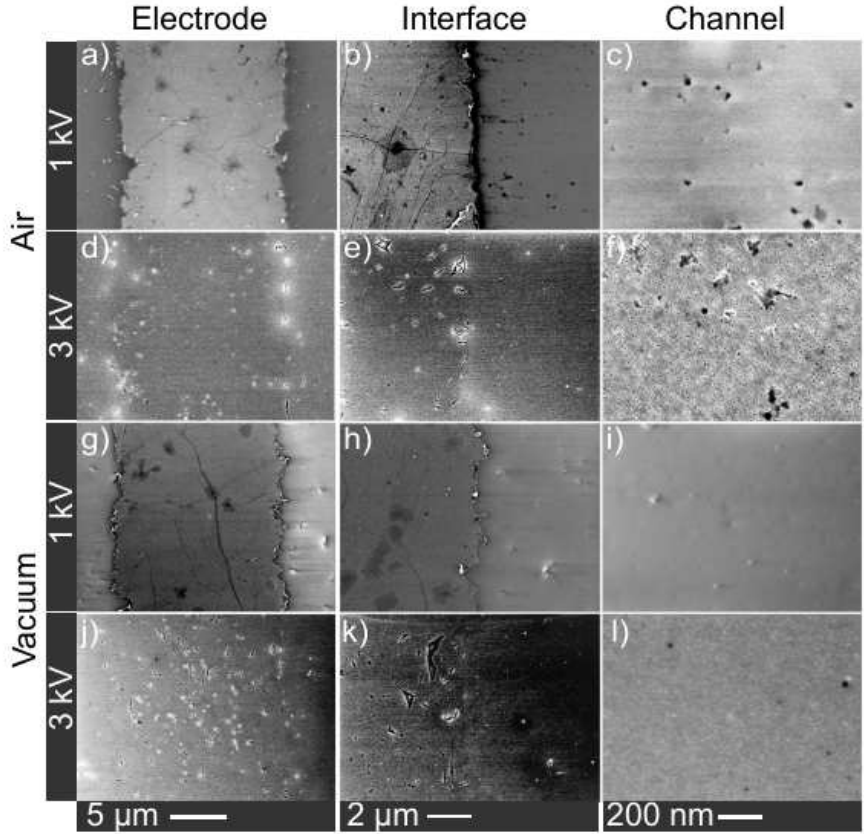}
\caption{Scanning electron microscopy images of graphene patterned in air (first two rows) and graphene patterned in vacuum (last two rows) show a reduction of debris on the sample, when laser patterning graphene in vacuum. Three regions of the transistor are analyzed: the graphene electrode (first column, marked by purple star in Fig.~\ref{setup}), the interface of the electrode to the channel (second column, green star in Fig.~\ref{setup}) and the channel region between source and drain electrodes (third column, yellow star in Fig.~\ref{setup}). The images of these regions are displayed after laser patterning without additional layers (a)-c) and g)-i)) and after TFT preparation with MO and passivation layer (d)-f)-j)-l)).}
\label{SEM}
\end{figure}
\subsection{Monolayer graphene patterning in air and in vacuum}
Ablation threshold of monolayer graphene (MLG) is determined by the well-known method of Liu \cite{Liu.1982}, the semilog plots of diameter square vs pulse energy can be found in the SI. The thresholds of the layer in air $F_{th,air}~=~74~\mbox{mJ/cm}^2$ and in vacuum $F_{th,vac}~=~77~\mbox{mJ/cm}^2$ are very similar, taking into account the power measurement accuracy of $\pm~5~\%$. These values are lower than the ablation threshold of the silicon substrate with thermally oxidized silicon dioxide ($F_{th,Si}~=~138~\mbox{mJ/cm}^2$).
For patterning functional structures with an appropriate electrode layout, the ablated areas (see Fig.~\ref{setup}) were scribed with a scanning speed of $v_{sc} = 0.2\,\mbox{m/s}$ and distance between lines of $\Delta y = 11\,\micro\mbox{m}$. This means a $95.7\,\%$ overlap of pulses within one line and $46.7\,\%$ overlap of lines is given, when taking into account the effective radius of $\omega_{eff} = 11.7\,\micro\mbox{m}$ calculated with Liu method \cite{Liu.1982}. 

The optimal ablation parameters were found by varying laser fluence, laser scanning speed and laser repetition rate. The fluence window below the damage threshold of the silicon substrate was evaluated in steps of $11~\mbox{mJ/cm}^2$ from $113~\mbox{mJ/cm}^2$ to $79~\mbox{mJ/cm}^2$. The scanning speed was varied from $6.1\,\mbox{m/s}$ to $0.2\,\mbox{m/s}$. Additionally, ablation experiments at different repetition rates in the range of $50\,\mbox{kHz}$ to $200\,\mbox{kHz}$, in steps of $50\,\mbox{kHz}$, were performed, while adapting the scanning speed to keep the same overlap of pulses of $95.7\,\%$. The laser ablation parameters, which resulted in the cleanest graphene edge quality, were found to be at a repetition rate of $200\,\mbox{kHz}$, fluence of $79~\mbox{mJ/cm}^2$ and a scanning speed of $0.2\,\mbox{m/s}$.

The topography of the patterned graphene electrodes was analyzed with SEM, see Fig. \ref{SEM}. For this purpose, three sections of the transistor are scanned: the patterned graphene electrode (first column), the interface between the electrode and the channel (second column) and the channel between the source and the drain electrodes (third column). These sections are marked by a purple, green and yellow star, respectively, in Fig. \ref{setup}. The first two rows of Fig. \ref{SEM} display images of MLG patterned in air and the last two rows contain images of MLG patterned in vacuum. In each case the first row displays the uncoated graphene after laser patterning and the second row displays the complete TFT, which consists of three layers: (1) MLG; (2) MO; (3) passivation layer. When comparing the SEM images it is clear, that the ablation of MLG in vacuum produces less debris than the ablation of MLG in air. Especially, when examining the debris in the channel between source and drain electrodes of the transistor. 
\begin{figure}[h]
\centering
\includegraphics[]{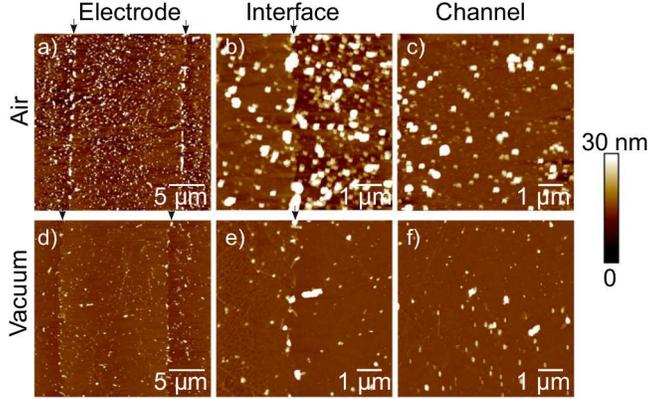}
\caption{Atomic force microscopy images of laser-patterned graphene electrodes a)-c) in air show higher occurrence of re-deposited material and formation of debris than the structures laser-patterned d)-f) in vacuum. The arrows mark the border of graphene and Si/SiO$_2$ substrate.}
\label{AFM}
\end{figure}

A reduction of debris formation while patterning MLG in vacuum is confirmed by AFM images, see Fig. \ref{AFM}. The AFM images were taken of the same three regions of interests as marked in Fig.~\ref{setup}: patterned graphene electrode, the interface between the graphene and the channel, and the channel region. In this case the patterned graphene sample is scanned without additional MO and passivation layers. In the channel region of the sample patterned in air (Fig. \ref{AFM} c)) a roughness of $R_{RMS}~=~6.02~\mbox{nm}$ with a maximum height of the profile of $R_{max}~=~121~\mbox{nm}$ was measured. The sample patterned in vacuum also displays redeposited material, but in this case a lower roughness of $R_{RMS}~=~2.22~\mbox{nm}$ with a maximum height of the profile of $R_{max}~=~53~\mbox{nm}$ in the channel region (Fig. \ref{AFM} f)).

\begin{figure}[h]
\centering
\includegraphics[]{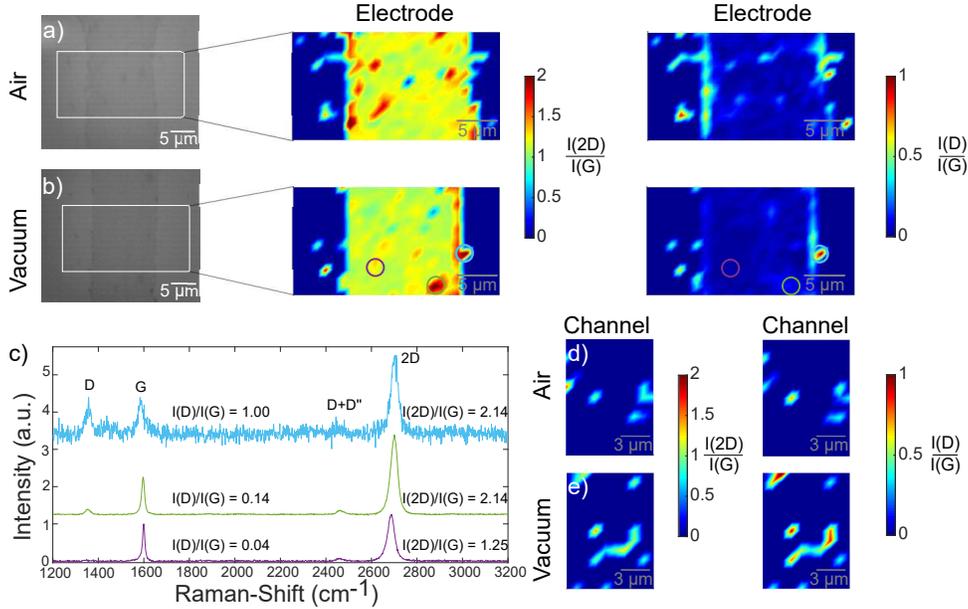}
\caption{Optical microscopy image and Raman map of I(2D)/I(G) ratio and I(D)/(G) ratio of the graphene electrodes laser patterned a) in air and b) in vacuum. c) Representative Raman spectra of the graphene electrode patterned in vacuum, normalized to the G peak with an vertical offset for visualization purposes. The corresponding positions in the Raman map are marked by circles with the matching colors in b). Raman map of I(2D)/I(G) ratio and I(D)/(G) ratio in the TFT channel region of samples patterned d) in air and e) in vacuum.}
\label{Raman_map}
\end{figure}

Raman mapping of laser patterned graphene shows that the intensity of the D peak ($1350~\mbox{cm}^{-1}$) is higher compared to the G peak ($1600~\mbox{cm}^{-1}$) at the edges of graphene electrodes, see Fig. \ref{Raman_map} a) and b). This is to be expected, as the D peak emerges due to defects in the carbon network and is more pronounced at the edges due to the disturbed translation symmetry of graphene \cite{Ferrari.2007}. Additionally, the D peak may also be attributed to oxidation of graphene edges, which occurs after breaking up the carbon bonds during laser ablation \cite{Kalita.2011,Zhang.2012b,Dong.2016}. 

The intensity of the 2D peak ($2700~\mbox{cm}^{-1}$) in the Raman spectra of graphene is very sensitive to the number of layers and its intensity is reported to be up to 4 times higher than the G peak intensity in MLG \cite{Ferrari.2006}. 
The mapping of the I(2D)/I(G) demonstrates that the peak ratio for the processed samples is approximately 2, i.e., somewhat lower to the values reported in the literature \cite{Ferrari.2006}. The left Raman maps of Fig. \ref{Raman_map} a) and b) show that the graphene electrodes display some spots with spectra where I(2D)/I(G) $\leq 1$. This signals may be from redeposited carbonaceous material on the graphene electrode itself after laser ablation.  Overall, the Raman maps show that the edges of the graphene electrode patterned in vacuum (Fig. \ref{Raman_map} b)) are smoother than the graphene electrode patterned in air (Fig.\ref{Raman_map} a)). 

Raman spectra taken in the channel region show that the debris and redeposited material observed before with SEM and AFM is attributed to carbonaceous material. These Raman spectra are displayed as a map in Fig. \ref{Raman_map} d)-e). The signal features high G and D peaks and a very low 2D peak. Therefore, the debris  produced during laser ablation is most probably carbon material that has been redeposited and oxidized during the laser ablation process.

\subsection{Thin film transistors with laser patterned graphene electrodes}
Prior to the preparation of the metal-oxide film the sheet resistance of the graphene electrodes was determined in four-point geometry. The sheet-resistance amounts $583\pm63~\Omega /\mbox{sq}$ and $542\pm 19~\Omega /\mbox{sq}$ for graphene electrodes patterned in air and in vacuum, respectively. Compared to the values specified by the manufacturer Graphenea ($440\pm 40~\Omega /\mbox{sq}$) the sheet-resistance is not significantly affected by the laser process, whereby laser processing under vacuum has a lower impact on the sheet resistance. This might be a consequence of suppressed oxidation of graphene during laser processing in vacuum.

\begin{figure}[h]
\centering
\includegraphics[]{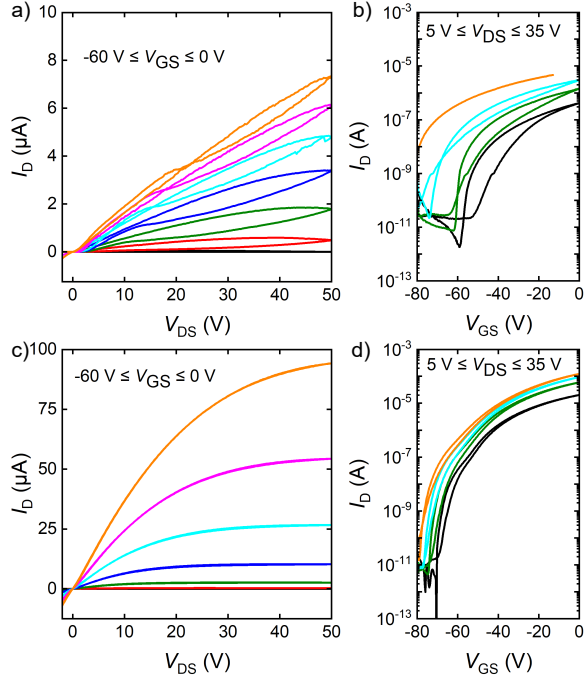}
\caption{Output and transfer characteristics of MOTFTs with graphene source and drain electrodes laser patterned in a)-b) air and c)-d) vacuum. The channel length (width) amounts $L~=~100~ \micro \mbox{m}$ ($W~=~ 500~ \micro \mbox{m}$).}
\label{TFTs}
\end{figure}

The output and transfer characteristics of the TFTs show a great difference in performance with source and drain MLG electrodes laser patterned in air compared to the MLG electrodes patterned in vacuum (Fig. \ref{TFTs}). The output characteristics of the transistor patterned in air shows hysteresis and does not saturate. This points to a parasitic conductivity in the channel. The transfer characteristics exhibit a clear shift to negative voltages during measurement. In contrast, the transistors with electrodes patterned in vacuum exhibit clear current saturation and a pinch-off behavior indicating that the entire thickness of the semiconductor channel layer can be depleted of free electrons (Fig. \ref{TFTs} c) and d)). More negative gate voltages are necessary to achieve full depletion of the active layer compared to TFTs with MLG electrodes patterned in air. A less pronounced shift to negative voltages is measured in the sample with graphene electrodes patterned in vacuum. Both samples have a high negative threshold voltage. The subthreshold swing $S_{sub}$ which is directly related to the interface trap density $N_S$ between the insulator and the MO, is extracted from the inverse of the maximum slope of the transfer characteristic. $S_{sub}$ of the TFTs with patterned graphene electrodes in air and in vacuum at $V_{DS}~=~5~\mbox{V}$ is around $2.5~\mbox{V/dec}$ and $1.5~\mbox{V/dec}$, respectively. This corresponds to a maximum interface trap density $N_S$ of $3.4\times 10^{12}~\mbox{cm}^{-2}$ and $2.0\times 10^{12}~\mbox{cm}^{-2}$ of the TFTs with graphene electrodes patterned in air and in vacuum, respectively. These values are higher compared to MOTFTs based on the same precursor with metal electrodes patterned by conventional lift-off technique ($S_{sub}~=~0.37~\mbox{V/dec}$) \cite{jaehnike.2015high}. We attribute this to a rough semiconductor/insulator interface arising from the debris. The roughly estimated field-effect mobility of $\micro _{FE}~\approx~0.1~\mbox{cm}^2\mbox{V}^{-1}\mbox{s}^{-1}$ and $\micro _{FE}~\approx~2.2~\mbox{cm}^2\mbox{V}^{-1}\mbox{s}^{-1}$ for TFTs with graphene electrodes patterned in air and vacuum, respectively, strongly underlines the relevance of a clean and smooth semiconductor/insulator interface. Compared to MOTFTs with conventional patterning methods and metallic electrodes $\micro _{FE}~\approx~28~\mbox{cm}^2\mbox{V}^{-1}\mbox{s}^{-1}$ the achieved field-effect mobility is still considerably smaller. Although a direct comparison is virtually impossible due to the large number of factors which must be taken into account, e.g. electrode material, passivation layer, interface treatment.

\section{Discussion}
 Laser-patterned graphene electrodes in vacuum, i.e. at reduced ambient pressure, exhibit less debris than the samples patterned in air. The size and the density of redeposited material depend on the ambient conditions due to two following pressure-dependent effects: reduction in the viscous drag force and in the density of the ambient gas. Both these effects reduce the kinetic energy dissipation of the ablated atoms, and hence enable them to fly far away from the sample surface. Measurements by Yoshida et al. demonstrated that the mean diameter of laser-ablated particles decreases with decreasing ambient gas pressure \cite{Yoshida1996}. As demonstrated by Geohegan \cite{Geohegan1992}, the cloud of ablated nanoparticles becomes denser and propagates slower at a higher ambient pressure. Thus, at a reduced pressure, the ablated particles are not confined in a vicinity of the sample surface but propagate further in the chamber and can even reach the chamber walls and be deposited there, as it happens in PLD (pulsed laser deposition) \cite{willmott2000}. 

The reduction of debris upon laser patterning of graphene in vacuum has a positive impact on the performance of MOTFTs compared to the transistors, with graphene electrodes processed in air. The high density of the redeposited particles in the channel of the sample laser ablated in air causes great perturbations in the thin MO layer. Since the deposition method of the MO is spin-coating, these particles lead to well known thickness variations and perturbations within the layer or cause the film not to be completely closed (see Fig. \ref{SEM} and \ref{AFM}). Thicker MO films result in the presence of a high-conductive back-channel layer at a distance beyond the screening length of the metal oxide/dielectric stack that induces humplike subthreshold transfer characteristics and more negative threshold voltages \cite{koo.2010,maeng.2011}. Additionally, the variations in film thickness might cause an incomplete conversion of the MO precursor, since the conversion parameters are optimized for a homogeneous layer \cite{weber.2014morph,jaehnike.2015high}. These effects lead to a parasitic conductivity causing the measured hysteresis and lacking saturation. Even though the density of particles per area is considerably decreased in the samples patterned in vacuum, these two effects can also occur to a smaller extent here. The negative threshold voltage and the high subthreshold swing indicate that a further reduction of debris is essential for high performance TFTs with laser patterned graphene electrodes. Once this is achieved the laser process is favorable compared to conventional patterning by UV lithography avoiding doping of the graphene electrodes due to resist residues \cite{chavarin.2016}.

\section{Conclusion}

Femtosecond laser patterning of CVD-grown graphene electrodes on silicon substrates and their application as source and drain electrodes in MOTFTs were studied. The quality of the patterned electrodes was improved by laser-ablation in vacuum due to debris reduction compared to laser processing in air which is consistent with previous reports on laser processing in vacuum \cite{chichkov.1996,matsumura.2005,feng.2015}. Raman spectroscopy analysis showed that the I(2D)/I(G) ratios for samples processed in air and in vacuum ($\approx 2$) are comparable, which suggests that having air or vacuum as environment during patterning does not considerably influence the graphene layer. A slight increase of I(D)/I(G) ratios on the edges of the patterned graphene electrodes with respect to the not laser-processed areas regardless of the ambient conditions indicates an increase of defects in these regions. Graphene laser patterning in vacuum leads to an improvement of MOTFT characteristics and a reduction of the semiconductor/insulator trap density.\\
Most notably, this is the first study to our knowledge to demonstrate MOTFTs using femtosecond laser-patterned graphene electrodes. However, some limitations are worth noting. Improving the graphene edge quality and a further reduction of debris, e.g. by enhancing the vacuum quality during laser processing is desirable. 

\section*{Acknowledgements}
We would like to thank the financial support of the Federal Ministry of Education and Research of Germany within the m-era.net project "CMOT-Investigation and tuning of graphene electrodes for solution-processable metal oxide thin-film transistors in the area of low-cost electronics" (03XP0014). We gratefully acknowledge the support by the Center for Interface-Dominated High Performance Materials (ZGH) at the Ruhr University Bochum for the access to the AFM.


\end{document}